\documentstyle[12pt,a4,epsf]{article}

\newcommand{\news}{\setcounter{equation}{0}}
\renewcommand{\theequation}{\thesection.\arabic{equation}}
\newcommand{\grad}{\mbox{$\bigtriangledown$}}
\begin{document}

\title{Ka\l u\.{z}a-Klein electrically charged black branes in M-theory}

\author{Miguel S. Costa\thanks{email address:
M.S.Costa@damtp.cam.ac.uk}\\and\\ Malcolm J. Perry\thanks{email address:
malcolm@damtp.cam.ac.uk}\\\\
D.A.M.T.P.\\
University of Cambridge\\
Silver Street\\
Cambridge CB3 9EW\\
England}

\date{October 1996}

\maketitle

\begin{abstract}
We present a class of Ka\l u\.{z}a-Klein electrically charged black
$p$-brane solutions of ten-dimensional, type IIA superstring
theory. Uplifting to eleven dimensions these 
solutions are studied in the context of M-theory. They can be
interpreted either as a $p+1$ extended object trapped around the
eleventh dimension along which momentum is flowing or as a boost of
the following backgrounds: the Schwarzschild black $(p+1)$-brane or
the product of the $(10-p)$-dimensional Euclidean Schwarzschild manifold
with the $(p+1)$-dimensional Minkowski spacetime.
\end{abstract}

\newpage

\section{Introduction}\

It has been conjectured that there exists a consistent quantum 
theory in eleven dimensions, M-theory, whose effective field 
theory limit is $D=11$ supergravity \cite{t,w}. Due to the presence of
a 4-form field strength in this supergravity theory, there are
supersymmetric electrically charged 2-brane \cite{ds} and
magnetically charged 5-brane \cite{gu} solutions which are believed to play a
crucial role in a precise formulation of M-theory. Double dimensional
reduction of the supermembrane action \cite{bst} from eleven to ten dimensions
gives rise to the type IIA superstring action \cite{d}. This is a
strong evidence for some relation between an underlying quantum theory
in eleven dimensions and type IIA superstring theory. In fact, Witten
\cite{w} has argued that the strong coupling
behaviour of type IIA superstring theory is given by the reduction on
$S^{1}$ of $D=11$ supergravity, with the compactification radius
increasing with string coupling. This fact leads to the conjecture
that compactification of M-theory on $S^{1}$ gives rise to type IIA
superstring theory. The $E_{8}\times E_{8}$ heterotic string is also
believed to arrive in the M-theory reduction on $\frac{S^{1}}{Z_{2}}$
\cite{hw}. Further, the type II dualities and the heterotic-type I
dualities are related to the compactification of the M-theory on
$T^{2}$ and on a $Z_{2}$ orbifold of $T^{2}$,
respectively \cite{hw,s,a}. Ten-dimensional type IIB and heterotic-type
I theories at arbitrary coupling correspond to a particular point of
the moduli space of the M-theory on the
previous manifolds. This gives a fairly unified picture for a
eleven-dimensional origin of the different superstring theories. In
the hope of learning more about this mysterious theory, it is
important to reinterpret the classical solutions to the effective
field theory limit of some superstring theories from an
eleven-dimensional perspective. The aim of this paper is to consider
such interpretation for the Ka\l u\.{z}a-Klein electrically charged
solutions of type IIA superstring theory. 

Ten-dimensional $N=2A$ supergravity is the effective field theory
limit of type IIA ten-dimensional superstring theory. The
bosonic fields are the metric $g$, the 2-form field ${\cal B}$ and
the dilaton $\phi$ all arising from the NS-NS 
sector, and  the 1-form field ${\cal A}$ and 3-form field ${\cal A}_{3}$
arising from the R-R sector. The 1, 2 and 3-form fields can
be used to construct charged black brane solutions of $N=2A$ 
supergravity \cite{hs}. In particular the 1-form field has been
used to construct electrically charged black hole and magnetically
charged black 6-brane solutions. The extreme solutions preserve 
one half of the maximal supersymmetry. These are particularly 
important because its semi-classical quantisation is believed 
to be exact \cite{k,klopp}.

We study Ka\l u\.{z}a-Klein electrically charged black $p$-brane
solutions of type IIA superstring theory in both ten and eleven
dimensions. In section two all possible cases of this class of
solutions are considered. The following section is concerned with the
eleven-dimensional picture. In section four we give some concluding
remarks. For completeness the solutions for Ka\l u\.{z}a-Klein electrically
charged black branes in $D$ dimensions with arbitrary coupling of 
the dilaton field to the Maxwell field strength are presented in an
appendix. Magnetic duality is used to generate magnetically charged black
brane solutions. 

\section{Ka\l u\.{z}a-Klein electrically charged black branes}\
\news

The effective action for the massless background bosonic fields
in type IIA superstring theory,
obtained by expanding in string loop and sigma model couplings,
is the $N=2A$ supergravity action
\begin{equation}
\begin{array}{rcl}
I_{IIA} & = & \frac{1}{2}\int d^{10}x\sqrt{-g}
        \left[e^{-2\phi}\left(R+4(\grad \phi)^{2}-\frac{1}{2.3!}{\cal
        H}^{2}\right)-
        \frac{1}{2.2!}{\cal F}^{2}-\frac{1}{2.4!}{\cal F}_{4}'^{2}\right]\\
\\
& + & \frac{1}{4}\int {\cal F}_{4}\wedge {\cal F}_{4}\wedge {\cal B},
\end{array}
\end{equation}
where ${\cal F}_{4}'=d{\cal A}_{3}+{\cal A}\wedge {\cal H}$, ${\cal
  F}_{4}=d{\cal A}_{3}$, ${\cal H}=d{\cal B}$, ${\cal F}=d{\cal A}$
and ${\cal A}$, ${\cal B}$ and ${\cal A}_{3}$ are 1, 2 and 3-form fields
respectively. After rescaling
the metric $g_{mn}\rightarrow e^{\frac{\phi}{2}}g_{mn}$
and setting ${\cal B}$ and ${\cal A}_{3}$ equal to zero we have in the
  Einstein frame 
\begin{equation}
I_{IIA}=\frac{1}{2}\int d^{10}x\sqrt{-g}\left[R
-\frac{1}{2}(\grad \phi)^{2}
-\frac{1}{2.2!}e^{\frac{3}{2}\phi}{\cal F}^{2}\right]. 
\end{equation}
A class of black $p$-brane solutions to this action is ($0\le p\le 6$)
\begin{equation}
\begin{array}{l}
ds^{2}=\left(1+\frac{\alpha}{r^{7-p}}\right)^{\frac{1}{8}}\left[
  -\frac{1-\left(\frac{r_{H}}
{r}\right)^{7-p}}{1+\frac{\alpha}{r^{7-p}}}dt^{2}+\frac{dr^{2}}
{1-\left(\frac{r_{H}}{r}\right)^{7-p}}+r^{2}d\Omega_{8-p}^{2}+dy^{s}dy_{s}\right] ,
\\\\
\phi(r)=\phi _{0}+\ln\left(1+\frac{\alpha}{r^{7-p}}\right)^{\frac{3}{4}},
\\\\
**{\cal F}=Qe^{-\frac{3}{2}\phi}(\epsilon_{8-p}\wedge\eta_{p}),
\end{array}
\end{equation}
where $s=1,...,p$ and $\epsilon_{8-p}$ and $\eta _{p}$ are the
volume forms on the unit $(8-p)$-sphere and $R^{p}_{y}$, respectively.
The solution for $p=0$ was first found 
by Gibbons and Maeda \cite{gm} (see also \cite{ghs}). We note that
according to the interpretation given by Papadopoulos and Townsend
\cite{pt} these objects remain 0-branes after compactifying along the
branes spatial dimensions. $\phi _{0}$ is the expectation value
of the dilaton field at infinity and determines the string coupling
there. The electric charge is defined by 
$Q=\frac{1}{V_{p}}\int_{\Sigma }e^{\frac{3}{2}\phi}*{\cal F}$, where
$\Sigma =S^{8-p}\times R^{p}_{y}$ is an
asymptotic spacelike hypersurface and $V_{p}$ the
volume of $R^{p}_{y}$. The ADM mass per unit of $p$-volume
\cite{lu} and the electric charge can be written in terms 
of the constants of integration $\alpha$ and $r_{H}$
\begin{equation}
\begin{array}{l}
\frac{M}{A_{8-p}}=\frac{8-p}{2}r_{H}^{7-p}+\frac{7-p}{2}\alpha,
\\\\
\left(\frac{Q_{0}}{(7-p)A_{8-p}}\right)^{2}=\alpha \left(\alpha
+r_{H}^{7-p}\right),
\end{array}
\end{equation}
where $Q_{0}=Qe^{-\frac{3}{4}\phi _{0}}$. It is possible to define a
scalar charge associated with the dilaton 
field. Noting that the kinetic term of the dilaton field in the action
(2.2) is given by $\psi -\psi _{0}=\frac{\phi -\phi _{0}}{2}$, this
charge is defined by the asymptotic behaviour of $\psi -\psi _{0}$
\begin{equation}
\psi -\psi _{0}\sim \frac{\Sigma}{(7-p)A_{8-p}}\frac{1}{r^{7-p}}.
\end{equation}
The dilaton field in (2.3) gives
\begin{equation}
\frac{\Sigma}{(7-p)A_{8-p}}=\frac{3}{8}\alpha.
\end{equation}
By changing to null coordinates it is easily seen that $r=r_{H}$
is a coordinate singularity of the metric in (2.3) (provided that
$r_{H}^{7-p}\neq -\alpha$ and $r_{H}\neq 0$). The scalar
curvature and the Ricci tensor, which are associated with the sources of the
gravitational field, as well as the Weyl tensor all diverge for $r=0$ and
$r^{7-p}=-\alpha $ in both the string and Einstein
metrics.\footnote{The solution (2.3) is defined for
$1+\frac{\alpha}{r^{7-p}}>0$. The region $1+\frac{\alpha}{r^{7-p}}<0$
could be obtained by taking the modulus of
$1+\frac{\alpha}{r^{7-p}}$. The corresponding background interpolates
between the singularities at $r=0$ and $r^{7-p}=-\alpha $. The $r<0$
region is not very interesting either.}

\subsection{Global Structure}\

In order to study the global structure we consider all possible
values of $\alpha$ and $r_{H}$ restricted to the physical
conditions $M, Q^{2}\ge 0$. Inverting (2.4) we have
\begin{equation}
\begin{array}{l}
\alpha=\frac{M}{A_{8-p}}\left[-1\pm \sqrt{1+\frac{8-p}{(7-p)^2}
\left(\frac{Q_{0}}{M}\right)^{2}}\right],
\\\\
r_{H}^{7-p}=\frac{M}{(8-p)A_{8-p}}\left[(9-p)\mp(7-p)
\sqrt{1+\frac{8-p}{(7-p)^2}\left(\frac{Q_{0}}{M}\right)^{2}}\right].
\end{array}
\end{equation}
The cases (i) and (ii) below correspond to the upper and lower
sign choices, respectively.\\
 
\begin{description}
\item[(i)] $\alpha \ge 0$ and $r_{H}^{7-p}\le \frac{2}{8-p}\frac{M}{A_{8-p}}\
\ \ \left(r_{H}^{7-p}\ge-\frac{7-p}{8-p}\alpha \right)$.\\

The equality $\alpha =0$, $r_{H}^{7-p}=\frac{2}{8-p}\frac{M}{A_{8-p}}$
corresponds to the Schwarzschild black $p$-brane. For
$\frac{2}{8-p}\frac{M}{A_{8-p}}>r_{H}^{7-p}>0$ we have 
$0<Q_{0}^{2}<4M^{2}$ and the Penrose
diagram is the same as in the usual four-dimensional Schwarzschild case.\\

For $r_{H}=0$ we have an extreme solution.
The metric in (2.3) becomes
\begin{equation}
ds^{2}=\left(1+\frac{\alpha}{r^{7-p}}\right)^{\frac{1}{8}}
\left[-\frac{dt^{2}}{1+\frac{\alpha}{r^{7-p}}}+dx^{i}dx_{i}+
  dy^{s}dy_{s}\right],
\end{equation}
where $i=1,...,9-p$ and $r^{2}=x^{i}x_{i}$. This metric has symmetry
$R\times E(p)\times SO(9-p)$, where $E(p)$ is the $p$-dimensional
Euclidean group. According to the terminology followed in \cite{pt}
these solutions should be interpreted as 0-branes but 
compactification along the $y$ directions is assumed. 
Provided that some supersymmetry is left unbroken quantum
corrections should vanish and this solution is therefore believed to
be exact.

We now check that one half of the maximal supersymmetry is broken. For
vanishing fermionic, dilatino, 2 and 3-form 
background fields, the supersymmetric transformations for the
fermionic and dilatino fields in $D=10$, $N=2A$ supergravity are
\begin{equation}
\begin{array}{c}
\delta\psi_{m}=D_{m}\epsilon+\frac{1}{64}e^{\frac{3}{4}\phi}
\left(\Gamma_{m}^{\ \ np}-14\delta_{m}^{n}\Gamma^{p}\right)
\Gamma^{10}\epsilon {\cal F}_{np},\\
\\
\delta\lambda=-\frac{1}{2\sqrt{2}}\left(D_{m}\phi\right)\Gamma^{m}
\Gamma^{10}\epsilon+\frac{3}{16\sqrt{2}}e^{\frac{3}{4}\phi}
\Gamma^{mn}\epsilon {\cal F}_{mn},
\end{array}
\end{equation}
where $\Gamma_{ab...c}=\Gamma_{[a}\Gamma_{b}...\Gamma_{c]}$, we
use the gamma matrices algebra $\{\Gamma_{a},\Gamma_{b}\}=2
\eta_{ab}$ compatible with a real representation and
$\Gamma^{10}=\Gamma^{0}...\Gamma^{9}$. As usual $m,n,...$
denote $10D$ world indices and $a,b,...$ $10D$ tangent space indices.
Our field configuration preserves some unbroken supersymmetry if
$\delta\psi_{m}=0$ and $\delta\lambda=0$. The Killing spinor
that solves this equation is
\begin{equation}
\epsilon=\left(1+\frac{\alpha}{r^{7-p}}\right)^{-\frac{7}{32}}\epsilon_{0},
\end{equation} 
with $\Gamma_{0}\Gamma^{10}\epsilon_{0}=\mp \epsilon_{0}$ and $\epsilon
_{0}$ a constant spinor. The $\mp$ sign choice corresponds to positive
or negative charge , respectively. Thus one half of the maximal
supersymmetry is broken. This solution saturates the Bogolmol'nyi
bound for the ADM mass per unit of $p$-volume $2M=|Q_{0}|$
$[17-19]$. As expected this result does not depend on $p$ as
it is derived from the supersymmetry algebra.

To understand the global structure consider the vector $v$
orthogonal to the $r=const.$ hypersurfaces. In components
$v^{m}=g^{mn}\partial _{n}r$ and $v^{2}=\left(1+\frac
{\alpha}{r^{7-p}}\right)^{-1/8}$. Thus $v^{2}>0$ for $r>0$ but $v^{2}
\rightarrow 0$ for $r\rightarrow 0$. This means that the
singular hypersurface $r=0$ is null, i.e. it coincides with the
horizon. In fact, $r=0$ is a Killing horizon in a limiting
sense. The Penrose diagram shown in figure 1(a) disagrees with
\cite{gm}, where a diagram corresponding to a naked singularity
at $r=0$ was drawn. Our result is consistent with the proposal of
supersymmetry being the cosmic censor \cite{klopp}.\\

For $r_{H}<0$ we have a naked singularity at $r=0$.\footnote{For
$r_{H}^{7-p}=-\frac{7-p}{8-p}\alpha$ we have a naked singularity with
$M=0$ and $\left(\frac{Q_{0}}{(7-p)A_{8-p}}\right)^{2}=
\frac{\alpha^{2}}{8-p}$.}\\

\begin{figure}[t]
\centerline{\epsfbox{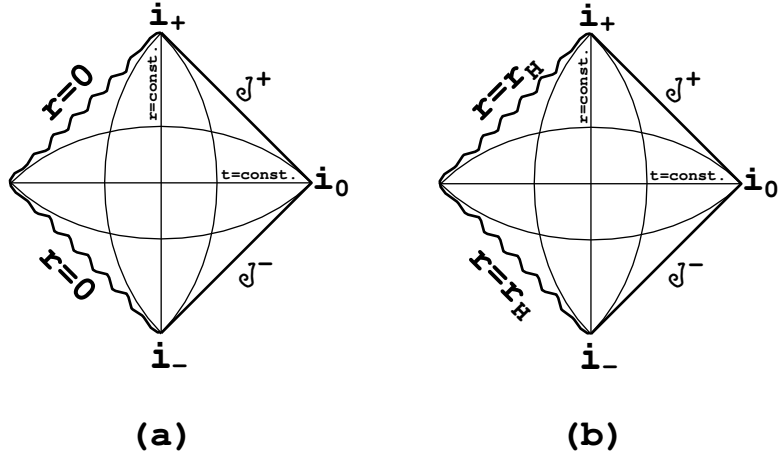}}
\caption{Penrose diagrams for (a) the extreme supersymmetric solution
  and (b) the uncharged dilatonic black p-brane. Each point is
  $S^{8-p}\times R^{p}$.}
\end{figure}

\item[(ii)] $\alpha \le -\frac{2M}{A_{8-p}}$ and $r_{H}^{7-p}\ge
\frac{2M}{A_{8-p}}\ \ \ \left(-\frac{7-p}{8-p}\alpha \le r_{H}^{7-p} \le
-\alpha \right)$.\\

For $r_{H}^{7-p}=-\alpha =\frac{2M}{A_{8-p}}$ the metric in (2.3)
simplifies to
\begin{equation}
ds^{2}=\left(1-\left(\frac{r_{H}}{r}\right)^{7-p}\right)^{\frac{1}{8}}
\left[ -dt^{2}+dy^{s}dy_{s}+\frac{dr^{2}}
{1-\left(\frac{r_{H}}{r}\right)^{7-p}}+r^{2}d\Omega_{8-p}^{2}\right],
\end{equation}
and it has symmetry $P(p+1)\times SO(9-p)$, where $P(p+1)$ is the
$(p+1)$-dimensional Poincar\'e group. This is an uncharged dilatonic
black p-brane with the singularity located at the horizon. 
The Penrose diagram is shown in figure 1(b). A naive calculation of
the Hawking temperature based on Euclidean continuation will give a
divergent result. Correspondingly, the area of the horizon (calculated
along a surface of $t=const.$) is zero. All supersymmetries are broken.\\

For $\alpha <-\frac{2M}{A_{8-p}}$ and $r_{H}^{7-p}>\frac{2M}{A_{8-p}}$
we have a naked singularity at $r=(-\alpha)^{1/(7-p)}$.$^{2}$\\

\end{description}

\subsection{Thermodynamics}\

The coupling of the dilaton field to the Maxwell field strength, $a$,
determines the global structure and thermal
properties. In the previous subsection the solutions of item (i) with
$r_{H}>0$ represent black $p$-brane solutions. Therefore the
standard thermodynamical arguments should apply.

Analytically continuing to Euclidean spacetime and
examining the behaviour of the metric in the vicinity of the horizon
$r=r_{H}$, the Hawking temperature is seen to be
\begin{equation}
T_{H}=\frac{7-p}{4\pi}r_{H}^{\ \ \frac{5-p}{2}}\left( r_{H}^{7-p}+\alpha
\right)^{-\frac{1}{2}}.
\end{equation}
In the limit $r_{H}\rightarrow 0$ we have
\begin{equation}
T_{H}\rightarrow \frac{7-p}{4\pi}r_{H}^{\ \ \frac{5-p}{2}}
\left(\frac{|Q|}{(7-p)A_{8-p}}\right)^{-\frac{1}{2}}.
\end{equation}
For $p=6$ $T_{H}$ diverges, for $p=5$ it converges to a finite number
and for $0\le p\le 4$ it converges to zero. In figure 2 the
isothermals in the $Q-M$ plot are shown. These plots were presented in
\cite{gm}. Here we present just those that correspond to the Ka\l
u\.{z}a-Klein electrically charged black $p$-branes that have an eleven
dimensional origin. In figure 3 we show the $T-Q$ and $T-M$ plots.

The charge of the black $p$-branes arises as a
central charge in a supersymmetric algebra. If quanta of charge
cannot be radiated, the evolution due to the Hawking evaporation
process is at constant charge. The black $p$-branes will decrease its
mass until the extreme limit is reached while the temperature converges
to infinity, a finite value or zero according to $p=6$, $p=5$ or $0\le
p\le 4$, respectively. In the case of $p=6$ we expect the
thermodynamic description to breakdown near the extremal limit. The
case of $p=5$ is rather puzzling.

\begin{figure}
\vbox{\vskip-1in\centerline{
\hskip0em\epsfxsize=1.0\hsize\epsfbox{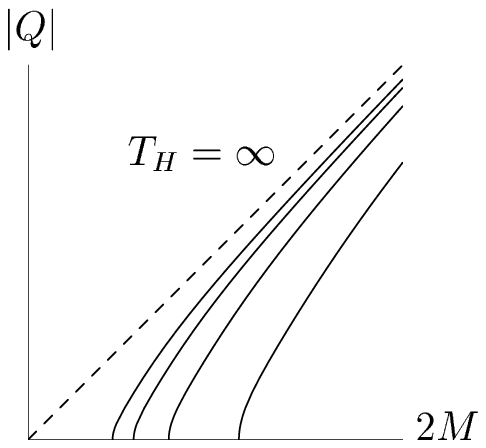}
\hskip-25em\epsfxsize=1.0\hsize\epsfbox{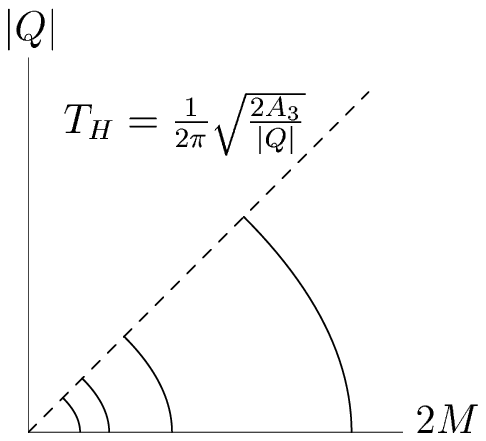}
\hskip-25em\epsfxsize=1.0\hsize\epsfbox{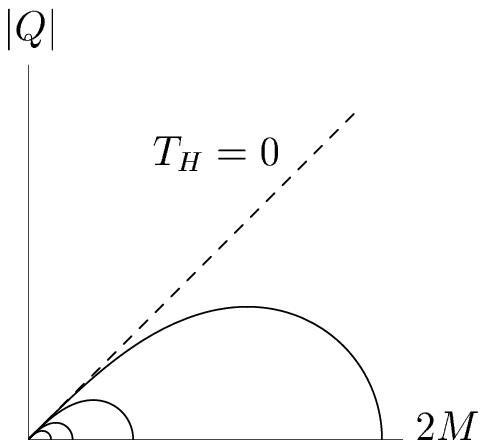}}
\vskip-5.7in}
\hbox{\hskip1.1in {$(a)$}\hskip1.5in {$(b)$}\hskip1.5in {$(c)$}\hfill}
\caption{Isothermals in the $Q-M$ plot for (a) $p=6$, (b) $p=5$
  and (c) $0\le p\le 4$. These cases correspond after
  dimensional reduction to black holes in $D=4$, $D=5$ and $6\le D\le
  10$ dimensions, respectively.}
\end{figure}
The equations (2.4) and (2.12) can be used to derive the mass formula
\begin{equation}
M=\frac{8-p}{7-p}T_{H}\frac{A_{H}}{4G}+\Phi _{H}Q,
\end{equation}
where in our units $4G=\frac{1}{2\pi}$, $A_{H}$ is the horizon area
(divided by $V_{p}$) and $\Phi _{H}$ is the horizon electric
potential. The differential form of (2.14) is the first law
of thermodynamics
\begin{equation}
dM=T_{H}\frac{dA_{H}}{4G}+\Phi _{H}dQ.
\end{equation}
The macroscopic specific entropy can be read from this
equation and it is given by
\begin{equation}
S_{BH}=2\pi A_{8-p}r_{H}^{\ \ \frac{9-p}{2}} \left( r_{H}^{7-p}+\alpha
\right)^{\frac{1}{2}}.
\end{equation}
\begin{figure}
\vbox{\vskip-1in\centerline{
\hskip0em\epsfxsize=1.0\hsize\epsfbox{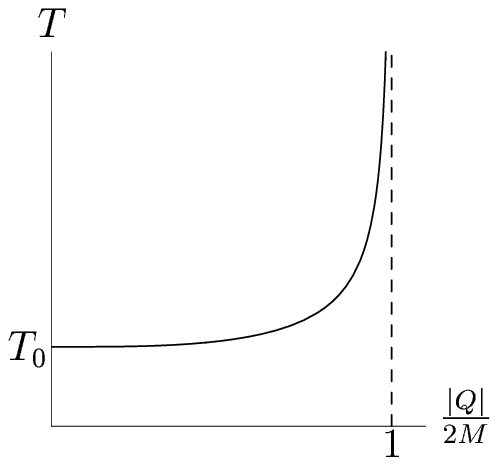}
\hskip-25em\epsfxsize=1.0\hsize\epsfbox{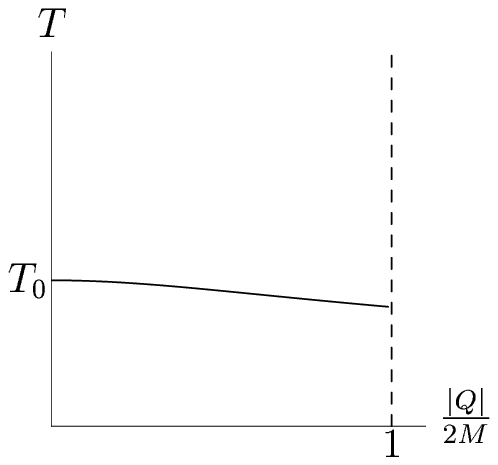}
\hskip-25em\epsfxsize=1.0\hsize\epsfbox{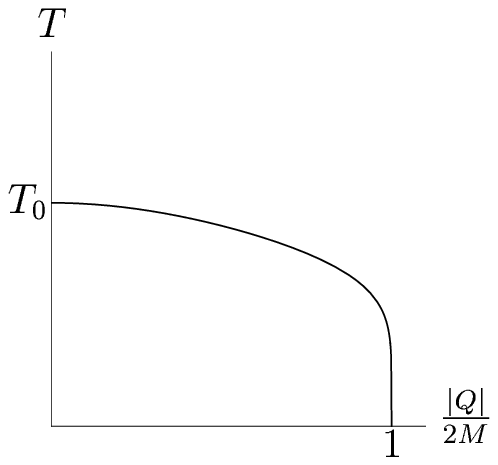}}
\vskip-5.8in}
\end{figure}
\begin{figure}
\vbox{\vskip-1in\centerline{
\hskip0em\epsfxsize=1.0\hsize\epsfbox{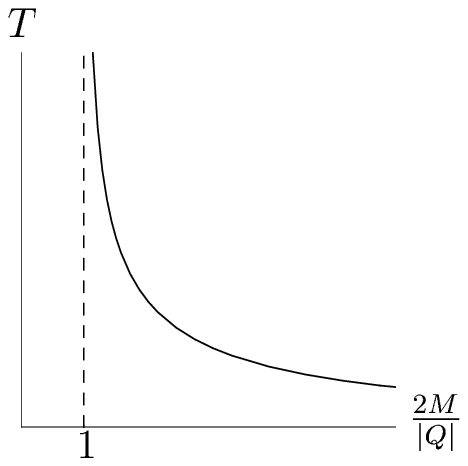}
\hskip-25em\epsfxsize=1.0\hsize\epsfbox{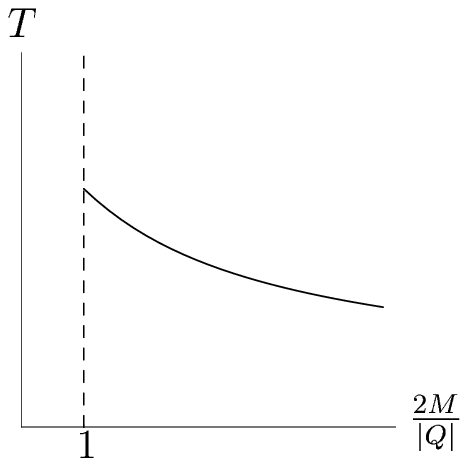}
\hskip-25em\epsfxsize=1.0\hsize\epsfbox{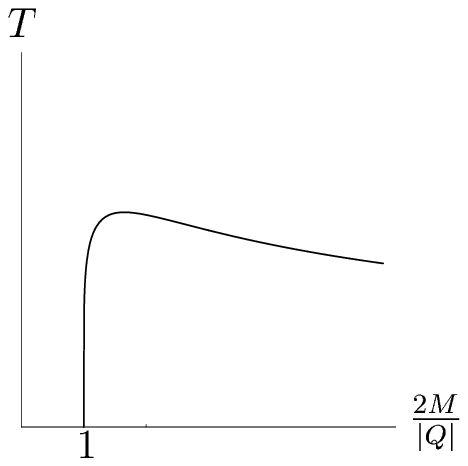}}
\vskip-5.7in}
\hbox{\hskip1.1in {$(a)$}\hskip1.5in {$(b)$}\hskip1.5in {$(c)$}\hfill}
\caption{$T-Q$ and $T-M$ plots for (a) $p=6$, (b) $p=5$
  and (c) $0\le p\le
  4$. $T_{0}$ depends on $p$ and is given by
  $T_{0}=\frac{7-p}{4\pi}\left(\frac{2M}{(8-p)A_{8-p}}\right)
  ^{-\frac{1}{7-p}}$. The specific heat $c=\frac{\partial M}{\partial
  T}|_{Q}$ is negative in cases (a) and (b) but changes sign in
  case (c) \cite{c}.}
\end{figure}

\section{Eleven-dimensional Spacetime}\
\news

Dimensional reduction of $D=11$ supergravity
on $M\times S^{1}$ yields $N=2$, non-chiral, ten-dimensional
supergravity whose bosonic sector action is given by (2.1)
\cite{cw}. The action for the bosonic sector of eleven dimensional
supergravity is
\begin{equation}
I_{11}=\frac{1}{2}\int
d^{11}x\sqrt{-g}\left[R-\frac{1}{2.4!}{\cal
    F}^{2}_{4}\right]+\frac{1}{12}\int
{\cal F}_{4}\wedge {\cal F}_{4}\wedge {\cal A}_{3},
\label{20}
\end{equation}
where ${\cal F}_{4}=d{\cal A}_{3}$ and ${\cal A}_{3}$ is a 3-form
field. Dimensional reduction to the Einstein frame is performed by
writing 
\begin{equation}
\begin{array}{c}
E_{M}^{A}=
\left(
\begin{array}{cc}
e^{-\frac{1}{12}\phi }e_{m}^{a} & -e^{\frac{2}{3}\phi }{\cal A}_{m} \\
0 & e^{\frac{2}{3}\phi }
\end{array}
\right)
\\
\\
{\cal A}_{MNP}={\cal A}_{mnp},\ {\cal A}_{MN10}={\cal B}_{mn},\ \
with\ M,N,P=0,...,9.
\end{array}
\label{21}
\end{equation}
Unless stated capital letters range from 0 to 10. $M,N,...$ denote $11D$
world indices and $A,B,...$ $11D$ tangent space indices.

The eleven dimensional metric corresponding to the
solution (2.3) is obtained in the following way
\begin{equation}
ds^{2}=g_{MN}dx^{M}dx^{N}=e^{-\frac{1}{6}\phi
}g_{mn}dx^{m}dx^{n}+e^{\frac{4}{3}\phi }\left(dx^{10}-{\cal
  A}_{m}dx^{m}\right)^{2},
\end{equation}
\begin{equation}
\begin{array}{ll}
ds^{2}= & -\left(1-\frac{r_{H}^{7-p}+\alpha}{r^{7-p}}\right)
dt^{2}+\left(1+\frac{\alpha}{r^{7-p}}\right)dx^
{\scriptscriptstyle{10}^{\scriptstyle{2}}} \mp
2\frac{\sqrt{\alpha \left(r_{H}^{7-p}+\alpha \right)}}{r^{7-p}}dtdx^{10}
\\\\  & +\frac{dr^{2}}{1-\left(\frac{r_{H}}{r}\right)^{7-p}} +
r^{2}d\Omega _{8-p}^{2}+dy^{s}dy_{s},
\end{array}
\end{equation}
where the $\mp$ sign choice corresponds to positive or negative
charge, respectively.\footnote{In the region $1+\frac 
{\alpha}{r^{7-p}}<0$ the first three terms in the metric have their
signs changed (see footnote 1).} Going from (3.3) to (3.4) we
performed the rescalings $x^{m}\rightarrow e^{\frac{1}{12}\phi
    _{0}}x^{m}$ and $x^{10}\rightarrow e^{-\frac{2}{3}\phi
  _{0}}x^{10}$.

The special cases $r_{H}=0$ and
$r_{H}^{7-p}=-\alpha $ will be treated separately afterwards. If
$\alpha \ge 0$ and $r_{H}>0$ it can be seen that $r=r_{H}$ is a
Killing horizon of $\frac{\partial}{\partial w}=\frac{\partial}
{\partial t}\pm \left(\frac{\alpha}{r_{H}^{7-p}+\alpha }\right)^{1/2}\frac
{\partial \ \ }{\partial x^{10}}$. This suggests that the metric
(3.4) can be written in the coordinates $w(t,x^{10})$ and
$y(t,x^{10})$ with $g\left(\frac{\partial}{\partial w},\frac
{\partial}{\partial y}\right)=0$. The transformation that does the job
is
\begin{equation}
\begin{array}{l}
w=\left(1+\frac{\alpha }{r_{H}^{7-p}}\right)t\mp
\frac{\sqrt{\alpha\left(r_{H}^{7-p}+
\alpha \right)}}{r_{H}^{7-p}}x^{10},\\
\\
y=\mp \frac{\alpha }{r_{H}^{7-p}}t+ \frac{\sqrt{\alpha\left(r_{H}^{7-p}+
\alpha \right)}}{r_{H}^{7-p}}x^{10},
\end{array}
\end{equation}
and the result is
\begin{equation}
ds^{2}=-\frac{1-\left(\frac{r_{H}}{r}\right)^{7-p}}{1+\frac{\alpha }
{r_{H}^{7-p}}}dw^{2}+\frac{r_{H}^{7-p}}{\alpha
}dy^{2}+\frac{dr^{2}}{1-\left(\frac{r_{H}}{r}\right)^{7-p}}+
r^{2}d\Omega _{8-p}^{2}+dy^{s}dy_{s}.
\end{equation}
After a rescaling of $w$ and $y$ we can see that $\alpha$ does
not play any role in the eleven-dimensional metric. From a $11D$
point of view it is just introduced by writing the metric in
the form (3.6) and changing to the $t,x^{10}$ coordinates
to obtain the straightforward compactifiable metric (3.4).
We now consider the different possibilities:\\

\begin{description}

\item[(i)] $\alpha \ge 0$, $r_{H}^{7-p}\le \frac{2}{8-p}\frac{M}{A_{8-p}}$
 and$\ r_{H}\neq 0\ \ \
\left(r_{H}^{7-p}\ge -\frac{7-p}{8-p}\alpha \right)$.\\

Consider first $r_{H}>0$. Performing the rescalings
$\sqrt{\frac{r_{H}^{7-p}}{r_{H}^{7-p}+\alpha }}w\rightarrow w$ and
$\sqrt{\frac{r_{H}^{7-p}}{\alpha}}y\rightarrow y$ we obtain the
Schwarzschild black $(p+1)$-brane metric
\begin{equation}
ds^{2}=-\left[ 1-\left(\frac{r_{H}}{r}\right)^{7-p}\right] dw^{2}+\frac{dr^{2}}
{1-\left(\frac{r_{H}}{r}\right)^{7-p}}+ r^{2}d\Omega
_{8-p}^{2}+dy^{2}+dy^{s}dy_{s}.
\end{equation}
The coordinate transformation (3.5) plus the $w$ and $y$
rescalings can be seen to generate a boost along the
compact direction with velocity given by 
$v^{2}=\frac{\alpha}{r_{H}^{7-p}+\alpha}$.
The Schwarzschild case $\alpha=0,\ r_{H}^{7-p}=
\frac{2}{8-p}\frac{M}{A_{8-p}}$ is obtained by setting $w=t$ and
$y=x^{10}$ in (3.7) (it corresponds to zero boost velocity,
i.e. no coordinate transformation).\\

For $r_{H}<0$ we perform the rescalings $\sqrt{\frac{-r_{H}^{7-p}}
{r_{H}^{7-p}+\alpha}}w\rightarrow w$ and $\sqrt
{\frac{-r_{H}^{7-p}}{\alpha}}y \rightarrow y$ and (3.6) becomes
\begin{equation}
ds^{2}=-dy^{2}+dy^{s}dy_{s}+
\left[1-\left(\frac{r_{H}}{r}\right)^{7-p}\right]dw^{2}
+\frac{dr^{2}}{1-\left(\frac{r_{H}}{r}\right)^{7-p}}+r^{2}d\Omega
_{8-p}^{2}.
\end{equation}
Now, the $y$ direction is timelike and the $w$ direction
spacelike. (3.5) plus the rescalings generate a boost along
the compact direction with velocity given by
$v^{2}=\frac{r_{H}^{7-p}+\alpha}{\alpha}$.
Because $r_{H}<0$ there is a naked singularity at $r=0$.\\

\item[(ii)]$\alpha <-\frac{2M}{A_{8-p}}$ and $r_{H}^{7}>\frac{2M}
{A_{8-p}}\ \ \ \left(-\frac{7-p}{8-p}\alpha \le r_{H}^{7-p}<-\alpha \right)$.\\

Performing the rescalings $\sqrt{\frac{r_{H}^{7-p}}
{-\left(r_{H}^{7-p}+\alpha \right)}}w \rightarrow w$ and
$\sqrt{\frac{r_{H}^{7-p}}
{-\alpha}}y \rightarrow y$ we obtain the metric (3.8)
but now $r_{H}$ is positive. As before the $y$ direction is timelike,
the $w$ direction spacelike and the boost velocity given by
$v^{2}=\frac{-(r_{H}^{7-p}+\alpha)}{-\alpha}$. The metric (3.8) (for
$p=0$) has been identified as a 
vacuum solution of $11D$ supergravity \cite{gu}. In the region $r_{H}<r<
(-\alpha)^{1/(7-p)}$, $m=\frac{\partial \ \ }{\partial x^{10}}$
becomes timelike and our compactification scheme breaks down as
can be seen in (3.3). This region is absent in the
solution (2.3).\footnote{To obtain the region
$1+\frac{\alpha}{r^{7-p}}<0$ we should have started with the
$10D$ metric defined in this region (see footnotes 1 and 3).}\\

\end{description}

Consider now the extreme cases $r_{H}^{7-p}=-\alpha$ and $r_{H}=0$:\\

\begin{description}

\item[(iii)]$r_{H}^{7-p}=-\alpha \ \ \ (\alpha<0)$.\\

The metric (3.4) becomes
\begin{equation}
ds^{2}=-dt^{2}+dy^{s}dy_{s}+
\left[1-\left(\frac{r_{H}}{r}\right)^{7-p}\right] 
dx^{\scriptscriptstyle{10}^{\scriptstyle{2}}}+
\frac{dr^{2}}{1-\left(\frac{r_{H}}{r}\right)^{7-p}}+r^{2}d\Omega
_{8-p}^{2}.
\end{equation}
This metric is similar to the one in (ii) if we replace $t$ by
$y$ and $x^{10}$ by $w$ (it corresponds to zero boost velocity,
i.e. no coordinate transformation). The spacetime is the product of
the $(10-p)$-dimensional Euclidean Schwarzschild manifold with the
$(p+1)$-dimensional Minkowski spacetime. If $x^{10}$ has period
$\frac{4\pi}{7-p}r_{H}$ we can avoid a conical singularity at
$r=r_{H}$ and our spacetime has a topology $R\times R^{2} \times S^{8-p}
\times R^{p}$. This solution has $P(p+1)\times SO(9-p)\times SO(2)$ 
symmetry. $x^{10}$ is
naturally periodic avoiding the singularity at $r=r_{H}$ that was
present in the ten dimensional version. At $r=r_{H}$ the
compactification radius vanishes in agreement with the fact that in
the ten-dimensional solution the dilaton field was $-\infty$
there. While in the latter the Hawking temperature was infinity and the
horizon area vanished, now $T=0$ because the time direction is flat
and $A(r=r_{H})=r_{H}^{8-p} A_{8-p} V_{p}$.\\

\item[(iv)]$r_{H}=0\ \ \ (\alpha>0)$.\\

The metric (3.4) simplifies to
\begin{equation}
ds^{2}=-dudv+\frac{\alpha}{r^{7-p}}du^{2}+dx^{i}dx_{i}+dy^{s}dy_{s},
\end{equation}
with $i=1,...,9-p$ and $v,u=t\pm x^{10}$ ($du^{2}\rightarrow
dv^{2}$ for negative charge). This is a plane
wave metric \cite{bko},
i.e. $\partial^{2}_{T}\left(\frac{\alpha}{r^{7-p}}\right)=0\
(r\neq 0)$, where $\partial^{2}_{T}$ is the Laplacian in the
transverse $x^{i}$ space. This solution has $SO(9-p)\times E(p)$ symmetry.

For vanishing fermionic and 3-form field backgrounds the
supersymmetric transformation for the fermionic field in eleven
dimensional supergravity is simply $\delta\psi_{M}=D_{M}\epsilon
$. The Killing spinor field solving $\delta\psi_{M}=0$ can be
related to $\epsilon ^{(10)}$ by
\begin{equation}
\epsilon^{(11)}=\left(1+\frac{\alpha}{r^{7-p}}\right)^{-\frac{1}{4}}
\epsilon_{0}=\left(1+\frac{\alpha}{r^{7-p}}\right)^{-\frac{1}{32}}
\epsilon^{(10)},
\end{equation}
with $\Gamma_{0}\Gamma_{10}\epsilon_{0}=\mp \epsilon_{0}$, $\epsilon
_{0}$ a constant spinor and the gamma matrices algebra
is now $\{\Gamma_{A},\Gamma_{B}\}=\eta_{AB}$.

The Ricci tensor for this solution can be written
distributionally as
\begin{equation}
R_{MN}=-2\partial^{2}_{T}\left(\frac{\alpha}{r^{7-p}}\right)U_{M}
U_{N}=2(7-p)\alpha A_{8-p}{\delta}^{9-p}(x^{i})U_{M}U_{N},
\end{equation}
with $U=\frac{\partial}{\partial v}$. Making use
of Einstein equations the energy-momentum tensor is
\begin{equation}
T^{MN}=2(7-p)\alpha
A_{8-p}{\delta}^{9-p}(x^{i})U^{M}U^{N}.
\end{equation}
This can be interpreted as a $p+1$ spatially extended object (or a string
according to \cite{pt}) sited at the
origin with a 11-velocity $U=\frac{\partial}{\partial v}$. It is
trapped around the eleventh dimension along which momentum is flowing.
Periodicity on $S^{1}$ of the wave function yields quantisation of
this momentum and therefore of the KK charge.\\

\end{description}

We could have done a similar analysis for the region
$1+\frac{\alpha}{r^{7-p}}<0$  and/or $r<0$ but the results are less
interesting. This exhausts all possible cases.

\section{Conclusion}\

From our results we can conclude that the class of ten-dimensional
black $p$-branes described by (2.3) when uplifted to eleven
dimensions on $M\times S^{1}$ have the following interpretation:
\begin{description}
\item[(i)] $\alpha \ge 0$, $r_{H}^{7-p} \le \frac{2}{8-p}\frac{M}{A_{8-p}}$
 and $\ r_{H}\neq 0$. If $r_{H}>0$ we have a boosted Schwarzschild
 black $(p+1)$-brane. The boosted and compactified direction is one of
 the brane's spatial directions. The
 ten-dimensional Schwarzschild p-brane is obtained when there is no
 boost. If $r_{H}<0$ we have a naked singularity.  
\item[(ii)]$\alpha
  <-\frac{2M}{A_{8-p}},\ r_{H}^{7-p}>\frac{2M}{A_{8-p}}$. The
  background is given by the boost of the product of the
  $(10-p)$-dimensional Euclidean Schwarzschild manifold with the
  $(p+1)$-dimensional Minkowski spacetime. The boosted and
  compactified direction is the Euclidean Schwarzschild time
  direction. The compactification scheme breaks down 
  for $r^{7-p}<-\alpha$ where $\frac{\partial}{\partial x^{10}}$
  becomes timelike. The corresponding ten-dimensional background has a
  naked singularity at $r^{7-p}=-\alpha$ where $\phi \rightarrow
  -\infty$, i.e. the string coupling vanish. 
\item[(iii)]$\alpha =-\frac{2M}{A_{8-p}}=-r_{H}^{7-p}$. Same
  background as in (ii) but without any boost. Provided that $x^{10}$
  has period $\frac{4\pi}{7-p}r_{H}$ the hypersurface $r=r_{H}$ is not
  singular. It corresponds in the ten-dimensional version  to a null
  singularity where the string coupling vanish.
\item[(iv)]$\alpha >0,\ r_{H}=0$. The solution corresponds to a $p+1$
  spatially extended object trapped around the eleventh dimension along
  which momentum is flowing. 
\end{description}

The latter is supersymmetric and its semiclassical quantisation
is therefore believed to be exact. This is consistent with one of the
key ideas of M-theory, namely solitons of string theory are elementary
excitations of M-theory.

Although these results were derived in the context of M-theory we
expect them to be generally valid for Ka\l u\.{z}a-Klein electrically
charged black branes in any spacetime dimension.

\section*{Acknowledgements}\

We acknowledge the financial support of JNICT (Portugal) under program
PRAXIS XXI (M.S.C.) and of DOE (M.J.P.).

\appendix

\section*{Appendix}\
\news
\renewcommand{\theequation}{A.\arabic{equation}}

In this paper we considered Ka\l u\.{z}a-Klein electrically charged black
branes in type IIA superstring theory. For completeness we will
present this solutions for the more general action
\begin{equation}
I=\frac{1}{2}\int d^{D}x\sqrt{-g}\left[R-\frac{1}{2}(\grad
\phi)^{2}-\frac{1}{2.2!}e^{a\phi}{\cal F}^{2}\right].
\end{equation}
The electrically charged black $p$-brane solutions are given by ($0\le
p\le D-4$)
\begin{equation}
\begin{array}{l}
ds^{2}=F_{\alpha}^{\ \delta}\left[-\frac{F_{H}}{F_{\alpha}^{\ \delta (D-2)}}dt^{2}+\frac{dr^{2}}{F_{H}}+r^{2}d\Omega
^{2}_{D-2-p}+dy^{s}dy_{s}\right],
\\\\
\phi =\phi _{0}+\frac{D-2}{2}a\ln F_{\alpha}^{\ \delta},
\\\\
**{\cal F}=Qe^{-a\phi}(\epsilon_{D-2-p}\wedge\eta_{p}),
\end{array}
\end{equation}
where $F_{\alpha}=1+\frac{\alpha}{r^{D-3-p}}$,
$F_{H}=1-\left(\frac{r_{H}}{r}\right)^{D-3-p}$, $s=1,...,p$ and 
$2\delta ^{-1}=D-3+\frac{D-2}{2}a^{2}$. The electric charge is defined
by $Q=\frac{1}{V_{p}}\int_{\Sigma }e^{a\phi}*{\cal F}$, 
where $\Sigma =S^{D-2-p}\times R^{p}_{y}$ is an asymptotic spacelike
hypersurface and $V_{p}$ the volume of $R^{p}_{y}$. The ADM mass per
unit of $p$-volume and the charge Q are given by
\begin{equation}
\begin{array}{l}
\frac{2M}{(D-3-p)A_{D-2-p}}=
\frac{D-2-p}{D-3-p}r_{H}^{D-3-p}+(D-2)\delta \alpha,
\\\\
\left(\frac{Q_{0}}{(D-3-p)A_{D-2-p}}\right)^{2}=(D-2)\delta \alpha \left(\alpha
+r_{H}^{D-3-p}\right),
\end{array}
\end{equation}
where $Q_{0}=Qe^{-\frac{a}{2}\phi _{0}}$. The Hawking temperature is
\begin{equation}
T_{H}=\frac{D-3-p}{4\pi}
\frac{1}{r_{H}\left(1+\frac{\alpha}{r_{H}^{D-3-p}}\right)
^{(D-2)\frac{\delta}{2}}},
\end{equation}
and the mass formula
\begin{equation}
M=\frac{D-2-p}{D-3-p}T_{H}\frac{A_{H}}{4G}+\Phi _{H}Q,
\end{equation}
where $\Phi _{H}$ is the horizon electric potential. The first
law of thermodynamics is still given by
\begin{equation}
dM=T_{H}\frac{dA_{H}}{4G}+\Phi _{H}dQ,
\end{equation}
and the macroscopic specific entropy is
 \begin{equation}
S=2\pi \left(1+\frac{\alpha }
{r_{H}^{D-3-p}}\right)^{(D-2)\frac{\delta}{2}}r_{H}^{D-2-p}A_{D-2-p}.
\end{equation}

Magnetic duality can be used to obtain magnetically charged black
$p$-brane solutions with the charge arising from a ($D-2$)-form field
strength. Let $\tilde{{\cal F}}=e^{a\phi}*{\cal
  F}=Q(\epsilon_{D-2-p}\wedge\eta_{p})$. If ${\cal F}$ is replaced by
$\tilde{{\cal F}}$ in the equations of motion that follow from (A.1)
we obtain a set of equations that can be derived from the action
\begin{equation}
\tilde{I}=\frac{1}{2}\int d^{D}x\sqrt{-g}\left[R-\frac{1}{2}(\grad
\phi)^{2}-\frac{1}{2.(D-2)!}e^{-a\phi}\tilde{{\cal F}}^{2}\right].
\end{equation}
Thus the magnetically charged black $p$-brane solutions are obtained from
the previous solutions by making the replacement $a\rightarrow -a$. The
magnetic charge is given by $Q=\frac{1}{V_{p}}\int_{\Sigma
  }\tilde{{\cal F}}$, with $\Sigma =S^{D-2-p}\times R^{p}_{y}$ an
asymptotic spacelike hypersurface.
 
In the cases where the actions I or \~I describe part of the bosonic
fields content of some supergravity theory we expect the extreme
solutions $r_{H}=0$ to be supersymmetric and to saturate the
Bogolmol'nyi bound for the ADM mass per unit of $p$-volume
$2M=\sqrt{\delta(D-2)}|Q_{0}|$.


\newpage

\begin{thebibliography}{22}
\bibitem{t}P.K.Townsend, Phys. Lett. {\bf B350} (1995) 184.
\bibitem{w}E.Witten, Nucl. Phys. {\bf B443} (1995) 85.
\bibitem{ds}M.J.Duff and K.S.Stelle, Phys. Lett. {\bf B253} (1991) 113.
\bibitem{gu}R.G\"{u}ven, Phys. Lett. {\bf B276} (1992) 49.
\bibitem{bst}E.Bergshoeff, E.Sezgin and P.K.Townsend, Phys. Lett. {\bf B189}
  (1987) 75; Ann. Phys. {\bf 185} (1988) 330.
\bibitem{d}M.J.Duff, P.S.Howe, T.Inami and K.S.Stelle,
Phys. Lett. {\bf B191} (1987) 70.
\bibitem{hw}P.Ho\v{r}ava and E.Witten, Nucl. Phys. {\bf B460} (1996)
  506; Nucl. Phys. {\bf B475} (1996) 94.
\bibitem{s}J.H.Schwarz, Phys. Lett. {\bf B360} (1995) 13,
  {\em Erratum} {\bf B364} (1995) 252; {\em Superstring dualities},
hep-th/9509148; Phys. Lett. {\bf B367} (1996) 97.
\bibitem{a}P.S.Aspinwall, {\em Some Relationships between dualities in
string theory}, hep-th/9508154.
\bibitem{hs}G.T.Horowitz and A.Strominger, Nucl.Phys. {\bf B360} (1991) 611.
\bibitem{k}R.Kallosh, Phys. Lett. {\bf B282} (1992) 80.
\bibitem{klopp}R.Kallosh, A.Linde, T.Ort\'{\i}n, A.Peet and
A.V.Proeyen, Phys. Rev. {\bf D46} (1992) 5278.
\bibitem{gm}G.W.Gibbons and K.Maeda, Nucl. Phys. {\bf B298} (1988) 741.
\bibitem{ghs}D.Garfinkle, G.Horowitz and A.Strominger, Phys. Rev. {\bf D43}
(1991) 3140, {\em Erratum} {\bf D45} (1992) 3888.
\bibitem{pt}G.Papadopoulos and P.K.Townsend, Phys. Lett. {\bf B380}
  (1996) 273.
\bibitem{lu}J.X.Lu, Phys. Lett. {\bf B313} (1993) 29.
\bibitem{w2}E.Witten, Commun. Math. Phys. {\bf 80} (1981) 381.
\bibitem{gh}G.W.Gibbons and C.M. Hull, Phys. Lett. {\bf 109B} (1982) 190.
\bibitem{dghr}A.Dabholkar, G.W.Gibbons, J.A.Harvey and F.R.Ruiz,
  Nucl. Phys. {\bf B340} (1990) 33.
\bibitem{c}B.Carter, Phys. Rev. Lett. {\bf 33} (1974) 558.
\bibitem{cw}I.C.G.Campbell and P.C.West, Nucl. Phys. {\bf B243} (1984)
112; F.Giani and M.Pernici, Phys. Rev. {\bf D30} (1984) 325; M.Huq and
M.A.Namazie, Class. Quant. Grav. {\bf 2} (1985) 293.
\bibitem{bko}E.A.Bergshoeff, R.Kallosh and T.Ort\'{\i}n,
  Phys. Rev. {\bf D47} (1993) 544.
\end{thebibliography}
\end{document}